# RF-PUF: IoT Security Enhancement through Authentication of Wireless Nodes using In-situ Machine Learning

Baibhab Chatterjee, Debayan Das and Shreyas Sen, *Senior Member, IEEE*
School of Electrical and Computer Engineering, Purdue University, West Lafayette, Indiana – 47907, USA
E-mail: {bchatte, das60, shreyas}@purdue.edu

*Abstract*—Physical unclonable functions (PUF) in silicon exploit die-to-die manufacturing variations during fabrication for uniquely identifying each die. Since it is practically a hard problem to recreate exact silicon features across dies, a PUF-based authentication system is robust, secure and cost-effective, as long as bias removal and error correction are taken into account. In this work, we utilize the effects of inherent process variation on analog and radio-frequency (RF) properties of multiple wireless transmitters (Tx) in a sensor network, and detect the features at the receiver (Rx) using a deep neural network based framework. The proposed mechanism/framework, called RF-PUF, harnesses already-existing RF communication hardware and does not require any additional PUF-generation circuitry in the Tx for practical implementation. Simulation results indicate that the RF-PUF framework can distinguish up to 10000 transmitters (with standard foundry defined variations for a 65 nm process, leading to non-idealities such as LO offset and I-Q imbalance) under varying channel conditions, with a probability of false detection < $10^{-3}$.

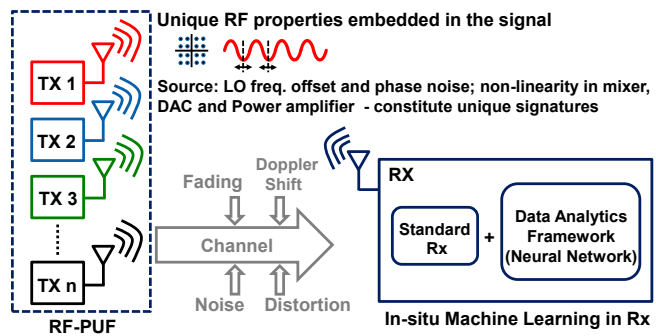

Fig. 1. Visualization of RF-PUF at a system level

## I. Introduction

The recent advancements in cheap computing, communication, wearable and implantable technologies have enabled a myriad of devices to be connected through the Internet-of-Things (IoT). Many of these devices continuously operate under mobile and untrusted environments, containing potentially malicious attackers. Hence, identifying and authenticating each transmitting device in such a network is an imposing problem for the receiving device. One of the traditional authentication protocols, called symmetric-key cryptography, involves placing a secret key in a non-volatile memory (NVM) or a battery-powered SRAM which can subsequently be used to create a digital signature or hash based encryption. However, this technique is vulnerable to key-hacking and suffers from significant area and power overhead. Because of these reasons, Physical Unclonable Functions (PUF), that exploit manufacturing process variations to generate a unique and device-specific identity for a physical system [1]-[5], are becoming increasingly popular. PUFs are simpler, smaller and more energy-efficient than NVM/SRAM-based solutions, and do not require anti-tamper mechanisms to detect invasive attacks.

Based on the number of challenge-response pairs (CRP) that the PUFs can handle, they are usually classified into strong and weak PUFs. Weak PUFs support a small number of CRPs which is linearly proportional to the hardware complexity of the PUF. Strong PUFs, conversely, support a large number of CRPs and hence can avoid polynomial time attacks. As a result, strong PUFs are usually employed in device authentication applications.

A PUF-based authentication protocol can enable easy and secure identification of devices in an IoT environment. When any narrowband Tx device modulates digital data, the resulting wave inherently contains device-specific unique analog/RF properties such as frequency offset and I-Q imbalance. These non-idealities are usually discarded/minimized at the receiver end as they are irrelevant in terms of the transmitted information. However, if those non-idealities are effectively harnessed using an in-situ machine learning framework at the Rx, the entropy information can be extracted for each of these transmitters, which leads to secure identification. Fig. 1presents the RF-PUF framework in presence of channel impairments, wherein every transmitter work as an instance of the PUF. Since <u>RF-PUF requires no extra hardware at the Tx</u>, it is even more interesting for an asymmetric network (such as the body area network) where the Tx needs to be low power and the Rx is a hub that can operate at a higher bandwidth and power. Such a framework can find multiple applications such as surveillance, forensic data collection and intrusion detection [6]-[7].

The remainder of the paper is organized as follows: Section II lists the recent works in the domain of RF authentication. Section III describes the proposed RF-PUF, while Section IV presents the performance parameters. Section V summarizes our contributions and lists future directions and improvements.

## II. Related Work

RF fingerprinting [8]-[11]enables automatic identification of wireless nodes in a network by extracting specific temporal and frequency-domain properties during power-on. Two fundamental techniques are usually adopted for RF fingerprinting – a transient method and a steady-state method. The transient method offers consistent and acceptable classification accuracy only when the start and the end of the transient can be reliably identified [7]. Moreover, a high oversampling clock (0.5-5 Gs/s, as shown in [9]-[10]) is required to properly analyse the transient properties. An alternative approach utilizes the steady-state properties of the Tx, extracted when the circuit transients are stabilized. However, the steady-state method requires fixed random channel access (RACH) preambles along with techniques such as spectral averaging or matched filtering [6]-[7], as the extracted properties depend on the transmitted bit-stream. The steady-state analysis method does not require sophisticated and high-power communication modules, as the steady-state properties can easily be identified. In this paper, we combine the concept of PUF with RF fingerprinting and develop a system architecture that utilizes the embedded RF properties of the Tx to authenticate wireless nodes using a machine learning framework in the Rx. To the best of our knowledge, this is the first work with a preamble-less steady state approach which trains the learning engine with multiple data streams and different channel conditions. RF-PUF utilizes the inherent non-idealities in the TX, and hence no additional hardware is required to generate the PUF identity, which is an important feature of the proposed system.

### III. RF-PUF: DEFINITION AND PROPERTIES

Although the PUF properties of the system originate from the manufacturing process variations of the transmitters, the identification of each node is done in the Rx sub-system which extracts the following features from the received signal:

#### A. Frequency features

Every transmitter in a non-FDM wireless network should ideally operate at a fixed carrier frequency. However, due to the inherent variations in the oscillator, each Tx will have its unique frequency offset. This offset/error works as a prime feature for device identification as shown in [7] and [12]. Different standards allow different limits for the frequency offset. For example, the frequency error must be within 25 ppm of the center frequency (2.412 GHz) for the IEEE 802.11b standard. For a normal distribution, this corresponds to a standard deviation ($\sigma$) of 20.1 kHz. An accurate and known reference clock in the Rx can calculate frequency offsets for multiple transmitters. Alternatively, a carrier synchronizer module (which already exists in a standard Rx for LO offset compensation) can be employed in the receiver that finds the frequency offset and compensates for it, as shown in our implementation in Fig. 2. The machine-learning sub-system identifies the Tx devices using this ppm offset as a prime feature.

#### B. I-Q features

The in-phase (I) and quadrature (Q) components of the Tx signal contains unique amplitude and phase mismatches for each Tx, which arises from the digital to analog conversion, power-amplifier (PA) back-off and gain. Because of compressive nonlinearity, the outer symbols in the constellation are affected more than the inner symbols. Hence it is necessary to extract amplitude and phase information for all symbols separately.

#### C. Channel features

The communication channel can also introduce time and frequency dependent variations in the form of attenuation, distortion and Doppler shift. To estimate and compensate these non-idealities, an automatic gain control (AGC) block, a Root-Raised Cosine (RRC) filter and a Doppler corrector is employed in the Rx as indicated in Fig. 2. The RRC filter, helps reducing inter-symbol interference (ISI), while the AGC module provides a measure of the channel attenuation to an Artificial Neural Network (ANN). Similarly, the Doppler corrector module estimates and corrects the amount of Doppler shift due to mobility of Tx and Rx in the network, and provides the information to the ANN for compensation.

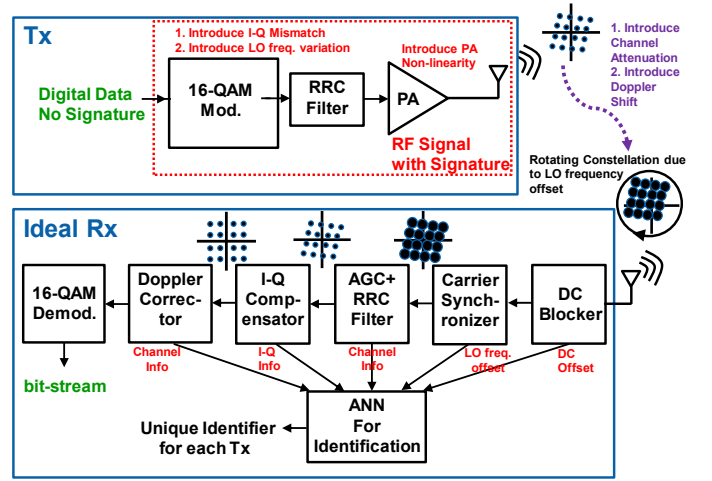

Fig. 2. Simulation setup involving Tx and Rx for RF-PUF implementation

#### D. 16-QAM system features

In modern transceivers, 16-QAM is employed to achieve a higher spectral efficiency. However, in our implementation, it also helps in extracting enough I-Q features (amplitude and phase imbalance) to identify the devices using the ANN.

#### E. Properties of the RF-PUF

The proposed system exhibits the properties of a strong PUF [3] that makes it suitable for authentication applications:

(a) *Constructability*: A PUF class P is constructible when a random PUF instance ($p_r \in P$) can be created by invoking a specific procedure, P.Create.

$$p_r \leftarrow \text{P.Create}$$

In case of RF-PUF, each transmitter is a PUF-instance. The technical limitations in the physical process of manufacturing the RF transmitters serve as the creation procedure $P.Create$ for the PUF instances.

(b) *Evaluability*: A constructible PUF class P is said to be evaluable when for a random PUF instance ($p_r \in P$) and a random challenge ($x$), it is possible to evaluate a response $y$.

$$y \leftarrow \text{P.Eval}[p_r(x)]$$

$x$ is the input bit-stream/challenge to each Tx (RF-PUF instance), and y is the unique response for Tx.

(c) *Reliability*: An evaluable PUF class P is reproducible/reliable if the intra-PUF-instance variation is probabilistically lower than a system-defined small number.

For any PUF, reproducibility is often represented as the worst-case difference ($D_{intra}$) between two distinct evaluations ($y(x), y'(x)$) on the challenge $x$ for a particular PUF instance.

$$D_{intra}(x) \quad dist[y_1(x), y_1'(x)]$$

$D_{intra}$ is the intra-PUF Hamming distance that serves as a metric to measure the resilience of the RF-PUF to varying environmental conditions (temperature, supply etc.). In an ideal scenario, $D_{intra} = 0\%$.

(d) *Uniqueness*: An evaluable PUF class P is unique if the inter-PUF-instance variation is probabilistically higher than a system-defined large number.

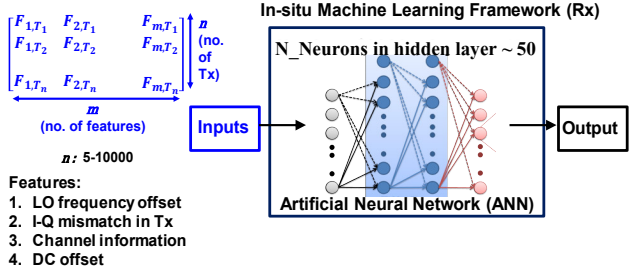

**In-situ Machine Learning Framework (Rx)**
N_Neurons in hidden layer ~ 50
Artificial Neural Network (ANN)

$m$ (no. of features)
$n$: 5-10000

Features:
1. LO frequency offset
2. I-Q mismatch in Tx
3. Channel information
4. DC offset

**Training Phase:**
- Train for multiple pseudo-random data streams in multiple iterations to enable preamble-less operation
- Include channel information for compensation

**Preamble-less Operation during Training (*utilized in this work*):**

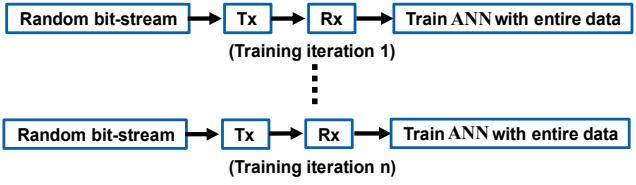

Random bit-stream → Tx → Rx → Train ANN with entire data
(Training iteration 1)
⋮
Random bit-stream → Tx → Rx → Train ANN with entire data
(Training iteration n)

**Goal: To obtain same accuracy as that of Preamble-based operation**

**Evaluation Phase:**
- Evaluate on the same $n$ Transmitters, with different channel conditions

Fig. 3. Preamble-less Training Framework for the ANN. This method enables on-the-fly authentication without knowing the expected bit-stream from the Tx.

The measure of uniqueness can be represented as the difference of the responses between two PUF instances $(y_1(x), y_2(x))$ for the same challenge, $x$.

$$D_{\text{inter}}(x) \quad dist[y_1(x), y_2(x)]$$

The quantitative results for uniqueness and reliability are evaluated in Section IV.

(e) *Identifiability*: A PUF class P is easily identifiable if it is reliable and unique, and if the intra-PUF variation is probabilistically lower than inter-PUF variation.

$$\text{Prob}(D_{\text{intra}}(x) < D_{\text{inter}}(x)) \approx 1$$

The simulation results shown in Section IV indicate that the worst case $D_{\text{inter}}$ is 3.9 ppm while the corresponding $D_{\text{intra}}$ is 2.9 ppm over 1000 Transmitters. This means that the instances are identifiable for 1000 or more devices.

(f) PUF Strength: RF-PUF can support a large number of challenge-response pairs (CRP), since we are using multiple analog/RF properties to define a single instance of RF-PUF. For $n$ distinct features in the real number space, the total number of CRPs for the RF-PUF will be $2^{16n}$, when the real-number space is represented by 16-bit numbers. Evidently, the number of CRPs is exponential in the number of features, and tends to a large number for even small values of $n$ (3-8). In other words, a probabilistic polynomial time (PPT) adversary can correctly predict the response to a new random RF-PUF instance/challenge with a negligible probability $(1/2^{16n})$. This confirms that RF-PUF is a "strong PUF" which is suitable for identification applications as shown in [1].

*F. Training the Neural Network*

To account for data variability in evaluation stage and to <u>enable preamble-less operation</u>, the neural network is trained in multiple iterations, and with different pseudo-random bit-streams (PRBS). The hyper-parameters, including the number of iterations required to compensate the effects of data variability were optimized aggressively. A feature matrix as shown in Fig. 3 was presented to the neural network as a training dataset. The extracted channel conditions for different training iterations were also presented as features which helps the network to learn to compensate for the channel variations. The number of hidden layers were also varied and was optimized for performance (detection accuracy). During Evaluation phase, a different pseudo-random data-stream from any of the transmitters is provided to the ANN for a different channel condition. Hence the neural network learns both data variabilities and channel variabilities simultaneously (at the cost of more training time) which lead to a robust system.

### IV. PERFORMANCE METRICS

The proposed PUF is simulated using the MATLAB Neural Network toolbox, and the End-to-End QAM simulation toolbox with RF impairments and corrections. To model the statistical manufacturing variabilities, foundry-provided process information of a standard 65 nm technology is included in the simulations. A total of 10,000 PUF instances were simulated in presence of varying channel conditions.

*A. Probability of False Detection*

Fig. 4(a) shows the accuracy/probability of false detection of a Tx device as a function of the total number of transmitters in the network, which indicates that the error is $< 10^{-3}$ for 4800 transmitters, and $< 10^{-2}$ for 10,000 transmitters. By increasing the number of neurons in the hidden layer from 10 to 50, the accuracy is improved. However, increasing the number of neurons to 100 causes overfitting, and does not improve the performance of the system even though the complexity and power cost of detection increases.

*B. Robustness to Channel*

The communication channel is always affected by free-space-path-loss, attenuation/noise in the medium (represented by $E_b/N_0$), interference, Doppler shift and fading. The contribution of $(E_b/N_0)$ has the most dominant effect as shown in [13]. We have analyzed and shown the effect of this channel noise on the probability of false detection in Fig. 4(b). Inter-symbol interference (ISI) leads to an increased error probability of $\approx 0.02$ in absence of the RRC filter when the mean $E_b/N_0$ of the channel is 15 dB, with a standard deviation of 10 dB. When the RRC filter is employed, ISI is reduced which makes the extraction of the I-Q features easier. This results in a reduced probability of error of $10^{-5}$ for the same amount if of $E_b/N_0$ variations as shown in Fig. 4(b).

In our implementation, the ANN is trained using variable data-streams in multiple iterations, which also includes variable amount of channel noise. As the number of training iterations increase, the performance of the system tends towards the performance achieved in fixed-preamble case. This overhead of having additional training iterations is simultaneously useful in terms of learning the channel variability, as each iteration will have a different channel condition which the network learns to compensate.

*C. Reliability (Intra-PUF Hamming Distance) and Uniqueness (Inter-PUF Hamming Distance)*

Fig. 4(c) presents the Reliability and Uniqueness of RF-PUF w.r.t. the input features provided to the ANN. Unlike a traditional digital PUF with multi-bit output, RF-PUF inherently embeds the unique signature of the Tx in the analog properties of the transmitted message, and hence the Intra-PUF and Inter-PUF distances can be plotted using the normalized parts-per-million (ppm) variation of the input features. However, a transformation of the feature spaces is required to represent the ppm variations of multiple features on a single axis. It should be noted that the total number of possible unique PUF instances is proportional to $\prod_{i=1}^{n} 2^{16}$ ($n$ = number of features, $2^{16}$= possible values for feature $i$ when represented with 16 bits), and hence we can intuitively utilize the geometric mean of the ppm values of all the features for a PUF instance. This geometric mean of the ppm values of the features (for 1000 transmitters) is used as the X-axis in Fig. 4(c).

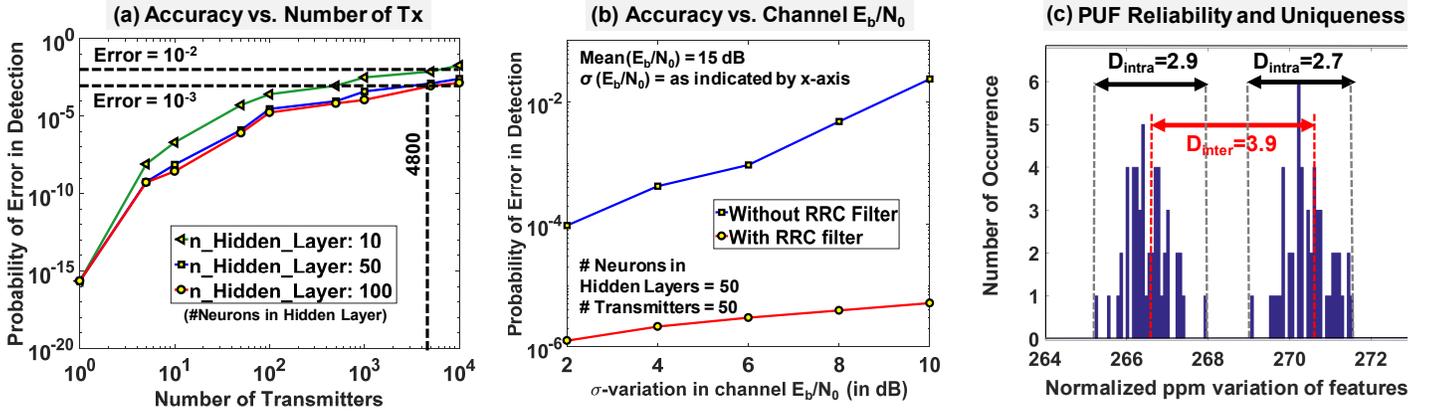

Fig. 4. (a) Probability of False detection as a function of the total number of transmitters in the system. (b) Probability of False detection as a function of the standard deviation ($\sigma$) of the $(\frac{E_b}{N_o})$ in the channel in presence of AWGN. (c) Worst case Intra and Inter-PUF distance with 1000 Transmitters and 50 hidden layers w.r.t. geometric mean of normalized ppm variation of features.

The worst case inter-PUF variation for 1000 transmitters is found to be 3.9 ppm, while the corresponding intra-PUF variation is 2.9 ppm. Hence, it is possible to uniquely identify the 1000 transmitters, but the probability of false detection keeps on increasing as the number of transmitters reaches a few thousands, as the difference between inter-PUF distance and intra-PUF distance reduces.

### D. Vulnerability to External Machine Learning Attacks

The RF-PUF employs supervised learning to uniquely identify the PUF instances. However, for an external attacker, the problem of correctly identifying a device in the network and masquerading itself as the same Tx device remains an unsupervised learning problem with a different learning framework. As it is not possible to list all the CRPs in polynomial time, a PPT adversary takes a very long time for the unsupervised learning.

Other PUF properties such as randomness (of the features), bias and stability with respect to temperature and supply is not a scope of this paper, and will be analyzed as a future work with hardware implementation where such notions can be properly justified.

## V. CONCLUSIONS AND FUTURE WORK

A conceptual development of RF-PUF is presented in this paper, supported by simulation results utilizing standard foundry-defined variations for a 65 nm technology. The inherent manufacturing non-idealities of a wireless Tx node can be exploited as a strong PUF for device identification in presence of channel imperfections and attenuation. An in-situ machine learning framework is shown to detect up to 10,000 devices with 99% accuracy. RF-PUF enables low-cost secure identification using intrinsic Tx properties embedded in the RF signal, without any extra hardware cost, and without any requirement for preamble-based or key-based authentication. The ANN in the Rx can be implemented on an application processor (on-board, already present in modern IoT nodes). As a future extension of this study, methods of compensating Rx non-idealities will be analyzed, along with efficient practical implementations using software defined radios. The randomness and stability of RF-PUF in presence of temperature and supply voltage variation will also be analyzed as part of the future work.

## ACKNOWLEDGEMENT

The work is supported in part by National Science Foundation (NSF) SaTC, CNS Grant No. 1719235 and Semiconductor Research Corporation (SRC) Grant No. 2720.001.